\documentstyle[11pt,preprint,aps,epsfig]{revtex}

\renewcommand{\baselinestretch}{1.0}   

\begin{document}
\title{Oscillatory Exchange Coupling across Cr$_{(1-x)}$V$_x$ Alloy
spacers}
\author{N. N. Lathiotakis$^\dag$ B. L. Gy{\"{o}}rffy$^\dag$,
E. Bruno$^\ddag$, B. Ginatempo$^\ddag$, S. S. P. Parkin$^\star$ }
\address{$^\dag$H. H. Wills Physics Laboratory, University of Bristol,
Tyndall Ave, BS8~1TL Bristol, U.K.}
\address{$^\ddag$Dipartimento di Fisica and Unit{\`{a}} INFM,
Universit{\`{a}} di Messina, Salita Sperone 31, 98166 Messina, Italy}
\address{$^\star$IBM Research Div, Almaden Res. Center, 650 Harry Rd,
San Jose, California~95120}
\date{\today}

\maketitle

\begin{abstract}
We have identified the pieces of the Fermi surface responsible for
the long period oscillations of magnetic coupling across Cr and 
Cr$_{(1-x)}$V$_x$ alloy spacers in metallic multilayers.
Analysing experiments and results of KKR-CPA calculations we find that
the periods are determined by the extremal wave~vectors of the hole
pockets centered on the N-point in the Brillouin zone.
\end{abstract}
\pacs{PACS: 75.70.-i,71.18.+y,71.23.-k,73.20.Dx}

\begin{figure}
\caption{\label{fig:3d}Schematic 3-D view of Cr Fermi surface.}
\centerline{\epsfig{figure=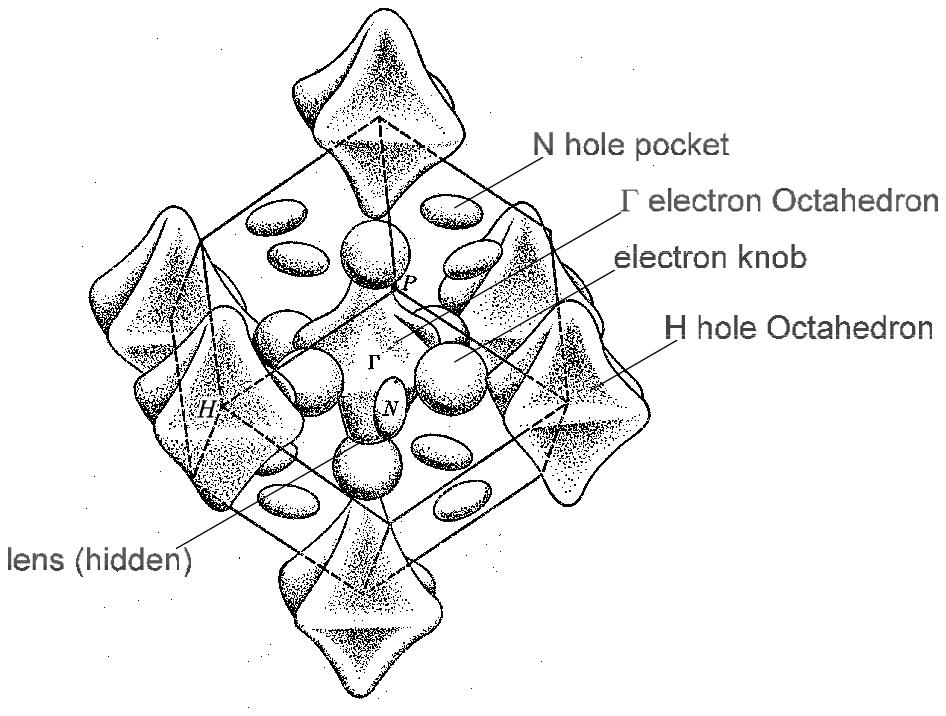,width=10cm}}
\end{figure}

{
The discovery that the magnetic coupling between two Fe layers separated 
by a Chromium layer, of thickness $L$(=1,100 \AA ), oscillates
from ferromagnetic to anti-ferromagnetic as function of $L$\cite{parkin3}
has lead to a veritable avalanche of observations of the same effect 
in similar metallic multi-layer systems\cite{heinrich}. The interest
in this surprising phenomenon derives from two sources. The first is its
relation to the technologically important GMR effect\cite{heinrich,baibich}.
The second is the wealth of evidence which links the periods of the
oscillations, $P_\nu$, to the Fermi surfaces of the bulk spacers\cite{heinrich}.
In this letter we shall be concern with the second aspect of the problem.

Whilst there are a variety of theoretical approaches for describing the 
above oscillatory magnetic coupling, each leads to the conclusion that
one is dealing with a variant of the well known RKKY interaction between
two impurity spins where the two point defects (spins) are replaced by two 
planar defects (the magnetic layers). Moreover, they all predict that the 
periods of oscillations, $P_\nu$, are determined by the extremal spanning 
wave-vectors, perpendicular to the layers, of the Fermi Surfaces of the bulk 
spacer metals\cite{baibich,edwards1}. Reassuringly, there are now  many cases 
where the calculated callipers vectors of the
appropriate Fermi surfaces are in good quantitative agreement with measurements
of the corresponding periods\cite{stiles}. Nevertheless, there is no consensus
as to which piece of the Chromium Fermi surface is responsible for the long
(18~\AA ) period of oscillation in the original, and since then much studied, 
experiments on the Fe/Cr/Fe 
system \cite{stiles,schilf_harr,koelling,dli,mirbt2,stiles2,tsetseris}. 
In what follows we shall present 
conclusive evidence that these oscillations are due to the hole pocket 
centered on the N-point in the bcc Brillouin zone  of bulk paramagnetic Cr.

Our arguments will be based on KKR-CPA calculations of the Fermi surfaces of
Cr$_{(1-x)}$V$_x$ alloys and a detailed, theoretical and experimental, 
investigation of the way the Fermi surface evolves as Vanadium is added 
to Cr. In short, we show that of all the possible candidate extremal vectors
only these spanning the  N hole-pockets change with concentration 
in a manner consistent with the experiments of Parkin et al\cite{schil2}.

A second, equally important point we wish to make, concerns the generic problem 
of the Fermi surfaces in random alloys of transition metals.  As has been 
emphasised repeatedly,\cite{beniamino} these fundamental features 
of the metallic state vary rapidly with changes in concentration and undergo
interesting electronic topological transformations\cite{beniamino}. 
Using the Cr-V system as an 
example, the first which features a non trivial topology, we shall demonstrate
that measurements of the oscillatory exchange coupling across an alloy can
be used as quantitative probe of subtle features of the Fermi surface. In fact, 
we suggest that they may provide comparable information to that extracted from
results of Angular Correlation of (positron) Annihilation Radiation (ACAR)
experiments\cite{alam}.

To be specific, let us recall the general theoretical 
result\cite{heinrich,edwards1,stiles} that
the exchange coupling between magnetic layers, across a pure metal spacer,
is given by
\begin{equation}
E= \frac{1}{L^2} \sum_\nu A_\nu \:\cos (\frac{2 \pi}{P_\nu}L + \phi_\nu)
\end{equation}
where $A_\nu$ is the amplitude, $\phi_\nu$ the phase and $P_\nu$ is the period 
of the $\nu$-th oscillatory contribution. As was shown in Ref.~\cite{schil2} 
and is confirmed by our KKR-CPA calculations, the states near the Fermi 
Energy in the Cr$_{(1-x)}$V$_x$ alloy case have sufficiently long lifetimes, 
as described by the Bloch Spectral Function (BSF) $A({\bf k},E)$, that the 
above formula can be used for these random alloys as well as for pure metals. 
Evidently, this means that spanning vectors are well defined by the sharp 
peaks in $A({\bf k},E)$. Taking this simplifying circumstance for granted we 
have carried out standard, self-consistent KKR-CPA calculations for a number of
concentrations $x$ and determined both the Fermi surfaces and their extremal
spanning vectors as described in Ref.~\cite{beniamino1}. All the theoretical 
results which feature in the foregoing discussion will refer to these  
calculations.

In fig.~\ref{fig:3d} we display a three dimensional, schematic picture of the
Fermi surface of Chromium. Clearly, for such a complex Fermi surface there 
are many extremal spanning vectors for each direction of the layer growth. 
Indeed, Stiles find no less than 6 candidates for explaining the 
long (18\AA ) period observed on (110) Fe/Cr/Fe multilayers and the 
controversy as to which of these, if any, is relevant continues 
unabated\cite{stiles,schilf_harr,koelling,dli,mirbt2,stiles2,tsetseris}.
 Of course,
a parameter free, reliable calculation of the amplitudes and the phases
of the oscillatory exchange coupling could differentiate among these 
candidates. However, such calculations, at the moment, are still very 
difficult\cite{stiles}. Model amplitude calculations have been done 
for Fe/Cr systems either using the reflection probabilities for 
Fe/Cr interface\cite{stiles2} or employing a Tight Binding model with
spin orbit interaction\cite{tsetseris}. Whilst they point towards the
conclusion  that the long oscillation period for pure Cr spacers originates 
from the N-centered hole pockets, they are not wholly convincing since the
size of the periods calculated agrees with the experiments only roughly.
In this letter we wish to advocate a much more effective procedure: Namely, 
we determine the periods $P_\nu$ as a function of the Vanadium concentration 
$x$ by the KKR-CPA calculations mentioned above and compare them with these 
deduced from measurements of the oscillatory coupling. As will be seen 
presently, only a small set of interrelated periods show the observed 
dependence on $x$.

In fig.~\ref{fig_pockets}a we show cuts across the Fermi surfaces of pure Cr 
and the Cr$_{0.85}$V$_{0.15}$ alloy parallel to the growth direction, (110),  
of the multi-layer system. Evidently, the `hole like' pieces of the Fermi 
surface increase and the `electron-like' ones decrease as $x$ changes from
$0$ to $0.15$. Consequently, some of the spanning vectors, identified by
arrows, increase or decrease with $x$ depending on whether they span electron 
or hole like pieces of the Fermi surface.  The behaviour of the corresponding 
periods $P_\nu$ with increasing V concentration is depicted in 
fig.~\ref{fig_pockets}b. Evidently, only the periods related to the N hole 
pockets decrease as in the experiments. Thus we conclude, in agreement with
the amplitude model calculations of Stiles\cite{stiles2} and 
Tsetseris {\it et al}\cite{tsetseris}, that it is these
N pockets that are responsible for the  observed long period.

To generate further evidence in support of the above conclusions all 
the different N-related
spanning vectors, which can be probed by measurements on samples with 
(110), (100) and (211) growth directions, have been studied. These 
investigations are summarized in fig.~\ref{fig_all_ellips}. The Fermi surface 
cross sections in figs.~\ref{fig_all_ellips}a and \ref{fig_all_ellips}b 
are well known\cite{stiles,schilf_harr,koelling,dli,mirbt2,tsetseris} and
correspond to paramagnetic Cr. Evidently N-pocket is not spherical for
pure Cr. According to our KKR-CPA calculations as V is added this pocket
shrinks without a change in shape or orientation. As a reflection of this
fact, the concentration dependences of the N-pocket related extremal wave 
vectors N$_1$, N$_2$ etc, shown in fig.~\ref{fig_all_ellips}c, are very 
similar.

Clearly, having eliminated the electron pieces of the Fermi surface, like
the `lens' or the `knob' as possible contenders for the role of causing
the long period oscillation the new issue becomes: the shape and orientation
of the N hole pocket. This is important since agreement between theory 
and experiments should establish the measurement of the oscillatory coupling
as novel and powerful probe of the Fermi surface. However such agreement 
will require further refinement of the experiments. A hopeful sign in this
direction is that although Fullerton {\it et al}\cite{fullerton} find the
oscillations to have the same periods, and even the same amplitude and phase
for all three of the growth directions (100), (110) and (211),
Tomaz {\it at al}\cite{tomaz}  reports small differences. Our predictions for 
the corresponding periods are N$_1$, N$_5$ and N$_9$ shown in 
fig.~\ref{fig_all_ellips}c. Clearly, whilst the differences are small, 
and hence the periods can be said to be roughly the same, 
in a more discerning experiment, they
should be observable.  

In conclusion, our calculations on the evolution of the Fermi 
surface of Cr$_{(1-x)}$V$_x$ with concentration identifies the 
N-hole pocket as the source of the long period oscillation of
the Exchange Coupling across Cr and Cr$_{(1-x)}$V$_x$ spacers.
The N-hole pocket is fairly isotropic but there are little differences
in the spanning vectors in different directions which could be 
measured by measuring the differences of the oscillation periods 
for different growth directions. Interestingly, the shape and 
orientation of the N-hole pockets can also be determined by 
2-dimensional 
ACAR experiments. Clearly, a quantitative agreement between the
KKR-CPA theory and the two above, very different, experiments
would be a significant step towards establishing the measurement 
of the above discussed periods as a quantitative probe of topologically
complex Fermi surfaces such as those which are host to
Electronic Topological Transitions\cite{beniamino} 
in transition metal alloys.\\
{\it Note Added:} After submitting the present letter for publication
we became aware of the work of D. Koelling \cite{koelling1} who has drawn
similar conclusion to us on the basis of the Rigid Band Model calculations.

One of the authors (NNL) was supported by the TMR network on 
`Interface Magnetism' (contract ERBFMRXCT960089) of the European Union. Also
 two of us (BG and EB) acknowledge financial support by INFM and the use of
the facilities at CECUM (Univ. of Messina).
}

\begin{figure}
\caption{\label{fig_pockets} (a) Cut of the pure Cr (solid)
and of Cr$_{0.85}$V$_{15}$ (dotted) Fermi surfaces, 
perpendicular to [1-1~0] direction
through $\Gamma$ point. 
(b) The Dependence of the oscillation periods (corresponding to the extremal
vectors shown in (a)) with V concentration. Experimental data from 
ref.~\protect\cite{schil2} for Fe/Cr$_{(1-x)}$V$_x$ and 
ref.~\protect\cite{parkin_co} for Co/V sandwiches are also included in (b) for 
comparison.  } 
\centerline{\epsfig{figure=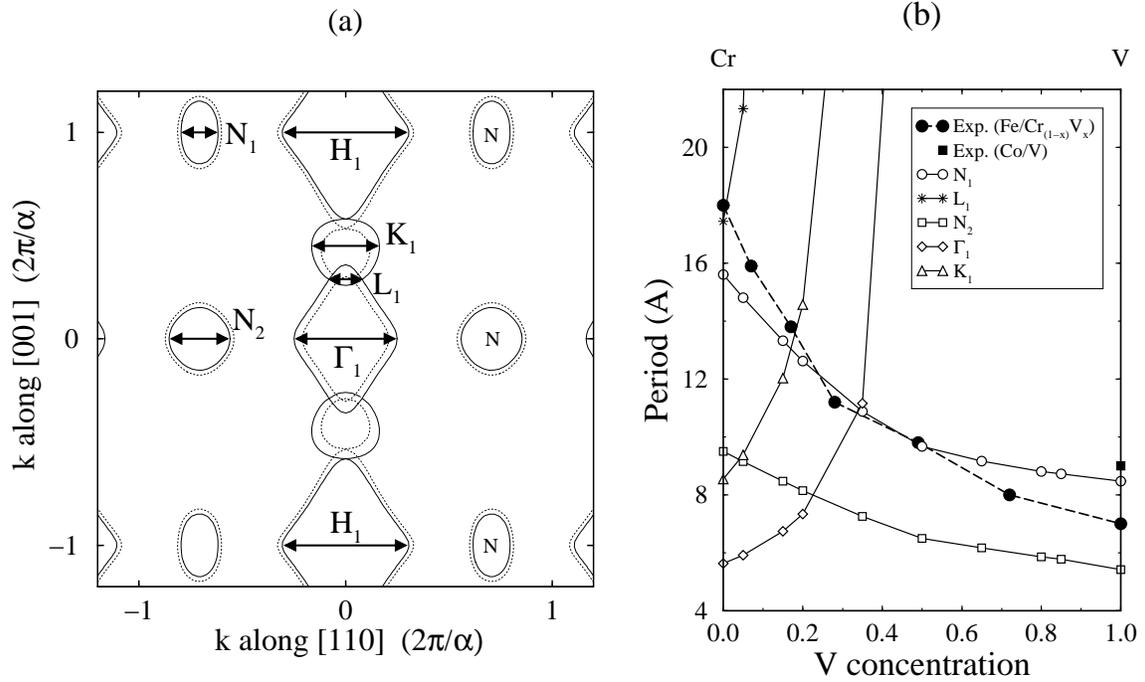,width=15cm}}
\end{figure}

\newpage
\begin{figure}
\caption{\label{fig_all_ellips}(a),(b) The Brillouin zone of BCC lattice with 
cuts of the Fermi volume by the two planes shown in the inset.
N$_1$, N$_2$, N$_3$ are the extremal vectors for the (110) direction,
while N$_4$, N$_5$ are those for the (100) and N$_6$, N$_7$, N$_8$, N$_9$
for the (211). The N$_4$ (not shown in (a) and (b) cuts) is the ellipsoid 
principal axis along the NP direction. N$_6$ and N$_8$ also are not shown 
in these cuts.
(c) The largest oscillation periods for each of the  (110), (100) and (211) 
directions corresponding to the  N$_1$, N$_5$ and N$_9$
extremal vectors respectively as functions of V concentration. 
The dependence on V concentration of the periods corresponding to the
rest of the extremal vectors is similar but the sizes of these periods are
significantly smaller than the ones plotted.
The experimental data of Parkin for Fe/Cr$_{(1-x)}$V$_x$~\protect\cite{schil2} 
and Co/V~\protect\cite{parkin_co} sandwiches 
are also included in (c) and refer to the (110) direction. 
}
\centerline{\epsfig{figure=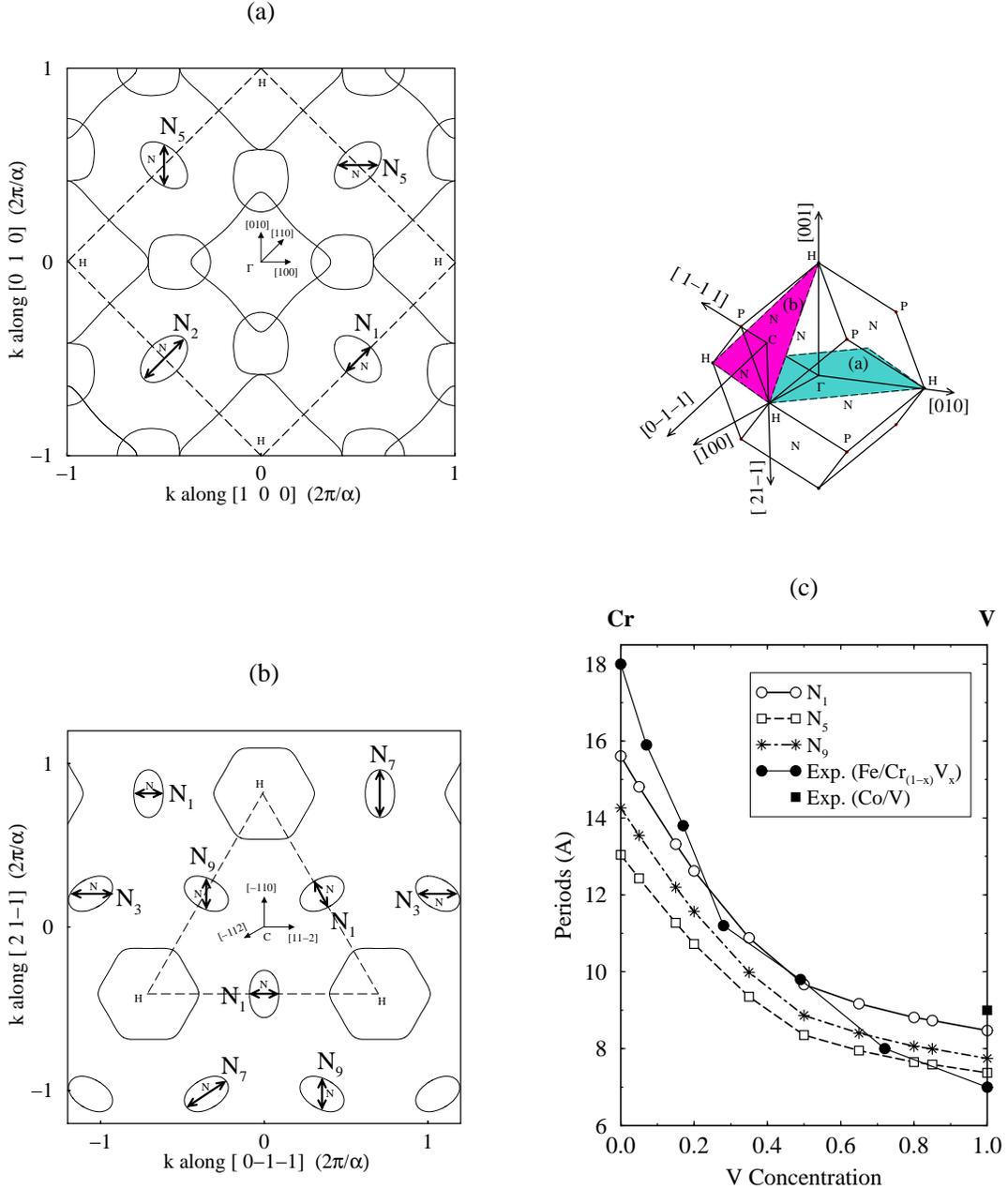,width=14cm}}
\end{figure}
\end{document}